\documentstyle[prl,aps,preprint]{revtex}
\tightenlines
\begin{document}
%\draft
%\twocolumn[{
\title{Josephson Junctions defined by a Nano-Plough.}
\draft
\author{B. Irmer, R.H. Blick, F. Simmel, W. G\"odel, H. Lorenz and 
J.P. Kotthaus}
\address{Center for NanoScience and Sektion Physik, 
Ludwig-Maximilians-Universit\"at M\"unchen,
Geschwister-Scholl-Platz 1, 80539 M\"unchen, Germany.}
%%%%%%%%%%%%%%%%%%%%%%% Kurzfassung %%%%%%%%%%%%%%%%%%%%%%%%
\date{\today}
\maketitle

\begin{abstract}
\widetext
\leftskip 54.8pt
\rightskip 54.8pt
We define superconducting constrictions by ploughing a deposited 
Aluminum
film with a scanning probe microscope.
The microscope tip is modified by electron beam deposition to
form a nano-plough of diamond-like hardness what allows the definition of highly transparent Josephson junctions.
Additionally a dc-SQUID is fabricated in order to verify the 
junction«s behaviour.
The devices are
easily integrated in mesoscopic devices as local radiation sources and
can be used as tunable on-chip millimeter wave sources.
\end{abstract}
%  }]
% %%%%%%%%%%%%%%%%%%%%%%%%%%%%%%%%%%%%%%%%%%%%%%%%%%%%%%%%%%%

\newpage

Ploughing is a well-known technique used since the earliest days of 
agricultural
cultivation~\cite{stone}.  By scaling this tool down in size to a few 
nanometers and combining it with
conventional scanning probe techniques, we facilitate ploughing of thin 
film
superconductors on semiconductor samples with nanometer resolution.
In the more common AFM scratching techniques, the tip is scanned
under strong loading forces to remove the substrate~\cite{hahn} or 
resist~\cite{sohn}.
Our technique utilizes the principle of ploughing in the same way as 
the traditional tool: 
material is removed from the substrate in a well-defined way, leaving 
behind deep trenches 
with the characteristic shape of the plough used.  

The advantages of applying a nano-plough for lithography are obviously 
the precision of alignment,
the non-damaging definition process compared to electron or ion beam 
structuring techniques, and 
the absence of additional processing steps, such as etching the substrate. 
Furthermore,
our technique allows the fabrication of highly transparent junctions
in comparison to the well-established
tunneling barriers commonly used to build Josephson devices. The main 
advantage of transparent
junctions is the lower resistance and hence the higher critical 
current densities $j_c$ of typically $10^{6}$A/cm$^2$
in comparison to $j_c\approx10^{2}$A/cm$^2$ for tunneling 
junctions~\cite{likharev}. 
This is expected to result in a dramatically enhanced microwave emission from these 
junctions. 
Recent theoretical work points out the importance of these junctions 
regarding
multiple Andreev reflection~\cite{zaitsev} and makes them of special 
interest in the field of mesoscopic physics.
One aspect in this field is the possibility of creating highly transparent Josephson junctions,
which radiate in the 
microwave range and which can be placed directly  
by the sample. 
Here we use the nano-plough to generate on-chip Josephson junctions 
suited as microwave sources.
We focus on the ploughing technique and on the Josephson 
characteristics of  
the constrictions formed. Finally we describe the superconduction junction«s 
characteristics and show data on single junctions and from the SQUID taken at dilution refrigerator 
temperatures. 

We define our devices in Aluminum thin films 
with thickness around 100~nm, which are thermally evaporated onto 
semi-insulating GaAs
substrates. In order to achieve reasonable superconducting properties, the 
background pressure 
is kept low, typically at $p\le 10^{-7}$~mbar with high evaporation 
rates $\approx$~100~\AA/s. 
2~$\mu$m wide wires and
loops with 10 $\mu$m inner radius are predefined with optical 
lithography. 

A crucial point for nano-ploughing is the appropriate plough: common 
scanning probe microscope
tips are either robust but too blunt to form small trenches 
(Si$_3$N$_4$ or hard coated tips)
or too brittle for this purpose (single crystal Si). We employ {\sl 
electron
beam deposition} (EBD) to grow material onto the end of common 
Si tips
by focusing the beam of a scanning electron microscope. 
The highly energetic electrons interact with additionally
introduced organic gases and form an organic compound, known as high 
dense carbon~\cite{wendel}. This material shows a hardness comparable to that of 
diamond~\cite{bert} in
combination with being non-brittle~\cite{nanotools}. 

For the nano-plough 500~nm long and 50~nm wide needle-like tips are 
grown under a non-zero angle
with respect to the axis of the pyramidal Si tip. Special care has to be taken to 
ensure a robust interface
between the EBD-tip and the silicon pyramid. When dragged through 
material with a lateral
force $F_l$, this angle causes an additional vertical force component 
$F_v$, which ensures
the plough to be pushed downwards, cutting its way through the
material (see upper inset in Fig. 1).\\
Imaging and positioning is done in standard AFM-mode. In the 
ploughing-mode, the tip is displaced by
nominally 1$\mu$m towards the surface, resulting in a loading force
$F_{\perp} \approx 40 \mu N$, sufficient to completely remove the
metal layer. Weak scratches in the GaAs substrate can be seen, thus 
avoiding a short-circuit
of the junction by a thin metal layer. To
define the Josephson constriciton of width $w$, the tip is withdrawn 
from the
surface, displaced by length $w$ and driven into the metal layer
again. The length $L$ of the Josephson contact is defined by the width
of the trench, which again is given by the tip diameter.
To date we were able to cut 300~nm thick Al films with a
minimum line width of 50~nm, yielding an aspect ratio of 1~:~6.
Fig. 1 depicts the SQUID geometry used with the optically predefined 
constrictions. The
ploughed lines can be seen perpendicular to the direction of the 
current flow.
The upper right inset depicts a schematic drawing of the nano-plough 
and the
force diagram as mentioned above. A typical junction is shown in the 
lower 
right inset with a width and length of $100 \times 100$~nm$^2$. The 
material removed
is pushed up onto the edges of the trench. 

In Fig.~2 the characteristic $IV$-trace of a single Al junction is 
shown, taken at a temperature of 35~mK.
Here, a current is fed through the junction and the voltage drop
across it is monitored (see inset with circuit diagram). The
dc-Jospehson current can be clearly seen. For currents larger than the
critical current $I_c$, a finite voltage drops across the junction,
defined by its normal resistance $R_n = 0.57~\Omega$. The curve agrees
well
with the theoretical description for a non-tunneling, Dayem-bridge like
superconducting weak link described by Likharev~\cite{likharev}.
The final devices show a critical
temperature of $T_c$ = 950 mK, compared to $T_c \approx$ 1.2~K for 
bulk Al. 
From its critical temperature we calculate the BCS gap 
$\Delta_{Al}(T\rightarrow 0) =
3.37 / 2\cdot k_B T_c = 138~\mu$eV \cite{BCS}.
The associated clean limit coherence lenght $\xi_{Al} = 0.18\cdot \hbar v_F / (k_B T_c)$
is then $\xi_{Al} = 2.9~\mu$m~\cite{tinkham}, 
using the bulk Fermi velocity of $v_F = 2.03 \times 10^6~$m/sec.
This should be compared to the
effective length $L_{eff}$ of the Josephson constriction, which is
somewhat larger than the 100~nm geometric length, but still an 
order of 
magnitude smaller than $\xi$, which makes our junction extremely 
transparent~\cite{likharev}\cite{zaitsev}. 

To be applicable as a radiation source, the critical
current and therefore the amplitude of the ac-current has to be large
and the capacitance has to be small so as not to shunt the junction. 
Both requirements are met by transparent junctions. The energy-gap 
frequency defines an upper limit for the emission at $\nu_{c} = 2\Delta / h \approx 
400$~ GhZ.
% very close to the intrinsic frequency $\omega_{c}/2\pi = 2e/h \cdot I_{c} \cdot R_{n} \cong 
%7.6 \times10^{11}$~Hz.
The critical current 
density (Fig.~2) is estimated to be $j_c = I_c / l\cdot d = 5.1 \times
10^{6}~A/cm^2$. From the temperature dependence of the critical 
voltage $V_{c}=I_{c}\cdot R_{n}$ (Fig.~3) we reproduce the universal, material 
independent
slope $\alpha = \partial(R_N \cdot I_c) / \partial T = 2\pi k_B / 7\zeta(3)e$ for $T 
\rightarrow
T_{c}$. This also identifies the conduction mechanism inside the 
junction: the Kulik-Omelyanchuk-theory (KO: straight 
line)~\cite{kulik} for 
the case of a clean weak link, i.e. $\xi,
l \gg L_{eff}$, where $l$ is the mean free path of electrons, fits
very well to our data. A tunneling junction in comparison would 
deliver much lower
critical currents (Ambegaokar-Baratoff: AB)\cite{ambegaokar}.

To verify the magnetic response of our junctions, we fabricated the 
two-junction SQUID shown in Fig.~1, where 
the maximum critical current oscillates as $I_{max} =
I_{c}\cdot{sin(\pi\phi/\phi_{0})}/{(\pi\phi/\phi_{0})}$, with $\phi$
being the trapped flux and $\phi_{0}$ being the flux quanta. The 
enclosed
area $A$ is (10$\mu$m)$^2\cdot\pi$, yielding an oscillation in
magnetic flux density of $\delta B = \phi_0 / A = (h/2e) / A = 6.4 
\mu$T,
which is well below the critical field of $B_{c}=15$~mT. In Fig.~4 the
critical current, taken from about 100 $IV$-traces, is plotted against
the applied external magnetic field. The SQUID junctions
had a width and length of $200 \times 100$~nm$^2$, therefore carrying
even higher critical currents than the single junction above. The 
measured period of 6.82~$\mu$T is in excellent agreement 
with the predicted $\delta B =  6.4 \mu$T.
The successful operation of the SQUID assures functioning 
of the Josephson junctions. Our current device is expected to 
generate radiation with frequencies around several 100~GHz, 
limited only by the capacitance of the junction and the pair breaking 
energy.\\

In summary, we have fabricated small weak-links in Al thin
films using a novel nano-plough technique. These junctions show 
high critical current
densities, low capacitive coupling and high intrinsic frequencies.
The junctions are ideally suited as tunable microwave on-chip devices.
A straightforward application of this new technique is the 
integration into mesoscopic systems, e.g. quantum dots or nanotubes. \\

% %%%%%%%%%%%%%%%%%%%%%% Danksagungen 
We would like to thank Armin Kriele and Stephan Manus
for their continuous support and the company NanoTOOLS GmbH for 
supplying AFM-tips.
We also thank W. Zwerger and R.J. Warburton for a critical reading of 
the manuscript.
We acknowledge A. Otto for help in archeology and
Lucy Brunner for support during the measurements.
The work was funded by the Volkswagen foundation under grant \# 
I/68769 and the Deutsche Forschungsgemeinschaft (SFB 348).

% %%%%%%%%%%%%%%%%%%%%%% Referenzen %%%%%%%%%%%%%%%%%%%%%%%%%
\vspace{0.5cm}
\small
{\it Electronic address: bernd.irmer@physik.uni-muenchen.de}\\

% %%%%%%%%%%%%%%%%%%%%%% Bilderverzeichnis %%%%%%%%%%%%%%%%%%

%%%%%%%%%%%%%%%%%%%%%%%%%%%%%%
\newpage
\begin{figure}[tb!]
{\caption{SEM micrograph of a double junction SQUID: The junctions are
generated in the Al film at the optically predefined 
constrictions.
The lower right inset shows a magnification of one of the junctions --
it can be clearly seen that all the material is removed.
The upper inset shows a schematic representation of
the nano-plough: the EBD tip is deposited with an angle on
a common AFM-tip, causing a vertical force $F_v$ when dragged through
material.
       }
       }
\end{figure}
%%%%%%%%%%%%%%%%%%%%%%%%%%%%%%%%%

\begin{figure}[tb!]
{\caption{Characteristic $IV$-trace of a single Josephson junction 
at T$_{bath}$=35~mK. The thickness of the Al
film is 100~nm, yielding a maximum critical current density of
$j_c = 5.1 \times 10^{6}$A/cm$^2$.
The upper inset shows the junction with a typical width of 100~nm, 
the lower inset gives a simplified circuit diagram.
        }
        }
\end{figure}
%%%%%%%%%%%%%%%%%%%%%%%%%%%%%%%%%

\begin{figure}[tb!]
{\caption{Temperature dependence of the normalized critical current~:
black dots indicate our measurements compared to the theory for a
tunneling junction
(Ambegaokar and Baratoff: AB) and for a clean weak link with
 L $\ll l, \xi$ (Kulik and Omelyanchuk: KO). KO theory predicts a 
critical current of $\pi$ for $T/T_{c}=0$.
The universal slope $\alpha = \partial(R_N \cdot I_c) / \partial T$ 
for
T $\rightarrow$ T$_{c}$ implies $\alpha = 615 \mu$V/K in 
good agreement with the theoretical value $\alpha = 635 \mu$V/K. This 
device shows a critical temperature T$_{c}$= 950~mK.
    }
        }
\end{figure}

\begin{figure}[tb!]
{\caption{SQUID-data showing the magnetic field dependence of the 
critical
current density. The line is a guide to the eye. The measured 
period of 6.82~$\mu$T is in excellent agreement 
with $\delta B = \phi_0 / A = (h/2e) / A = 6.4 \mu$T.
       }
       }
\end{figure}
%%%%%%%%%%%%%%%%%%%%%%%%%%%%%%


\begin{thebibliography}{10}
\bibitem{stone} To the best of our knowledge ploughs are depicted for the first 
time 
on small clay plaques found at Uruk IV in Iraq, dated around 3200 BC.
\end{thebibliography}
\end{document}